\documentclass[hyper,cits]{PoS}
\pdfoutput=1

\usepackage{amsmath}
\usepackage{amsxtra}
\usepackage{amssymb}
\usepackage{amsfonts}

\usepackage[american]{babel}
\usepackage[utf8x]{inputenc}
\usepackage{ucs}
\usepackage[T1]{fontenc}

\usepackage{feynmp}
\usepackage{array}
\usepackage{float}

\ifpdf
\DeclareGraphicsExtensions{.1}
\DeclareGraphicsExtensions{.2}
\DeclareGraphicsExtensions{.3}
\DeclareGraphicsExtensions{.pdf}
\else
\DeclareGraphicsExtensions{.pdf} 
\fi

\newcommand{\eref}[1]{Eq.~(\ref{#1})}

\newcommand{\hif}{\ensuremath{{F}}}
\newcommand{\qhifmunu}[1]
{\ensuremath{\frac{1}{16\pi^2}\operatorname{tr}%
\sum_{\mu,\nu}%
\left[\hif_{\mu\nu}#1\widetilde{\hif}_{\mu\nu}#1\right]}}

\usepackage[square,comma]{natbib}
\leftline{\parbox{3cm}{\large\rm HU-EP-07/42}
\vspace{1cm}}

\title{Smearing and filtering methods in lattice QCD --- %
a~quantitative comparison %
   \thanks{This work was supported by DFG under contract FOR 465. %
   \endgraf We like to thank the Leibniz Rechenzentrum in Munich for % 
   support and training. } %
 } 
\ShortTitle{Comparison of filtering methods}

\author{\speaker{Stefan Solbrig}\thanks{supported by DFG and BMBF}\\
Universit\"at Regensburg, Institut für Physik, 
93040 Regensburg, Germany\\
E-mail: \email{stefan.solbrig@physik.uni-regensburg.de}}

\author{Falk Bruckmann\\ 
Universit\"at Regensburg, Institut für Physik, 
93040 Regensburg, Germany\\
E-mail: \email{falk.bruckmann@physik.uni-regensburg.de}}

\author{Christof Gattringer\\
Universit\"at Graz, Institut für Physik, 8010 Graz, Austria\\
E-mail: \email{christof.gattringer@uni-graz.at}}

\author{Ernst-Michael~Ilgenfritz\\
Humboldt-Universit\"at zu Berlin, Institut f\"ur Physik, Newtonstrasse 15,
12489 Berlin, Germany\\ 
E-mail: \email{ilgenfri@physik.hu-berlin.de}}

\author{Michael~M\"uller-Preussker\\
Humboldt-Universit\"at zu Berlin, Institut f\"ur Physik, Newtonstrasse 15,
12489 Berlin, Germany\\ 
E-mail: \email{mmp@physik.hu-berlin.de}}

\author{Andreas Sch\"afer\\
Universit\"at Regensburg, Institut für Physik, 
93040 Regensburg, Germany\\
E-mail: \email{andreas.schaefer@physik.uni-regensburg.de}}

\abstract{We systematically compare three filtering methods used 
to extract  topological excitations from lattice gauge configurations,
namely smearing, Laplace filtering and the filtered fermionic topological
charge (with chirally improved fermions). 
Each of these techniques introduces ambiguities, like its parameter dependence. 
We show, however, that all these methods can be tuned to each other 
over a broad range of filtering levels and that they reveal very similar 
topological structures. For these common structures we find an interesting
power-law relating the number and packing fraction of clusters of filtered
topological charge.}

\FullConference{The XXV International Symposium on Lattice Field Theory\\
                 July 30 - August 4, 2007\\
                 Regensburg, Germany}

\begin{document}

\bibliographystyle{latprocutphys}

\section{Introduction}

\noindent Topological objects are known to be linked to chiral symmetry 
breaking. They may also have a close connection with confinement. Therefore, 
the study on the lattice of how the topological charge is distributed in 
space-time will deepen the insight in those mechanisms. However, there is no 
unique topological charge density on the lattice. Various methods to define it 
on lattice gauge configurations exist, but they need not to agree on a given 
configuration. Filtering techniques are necessary to extract a smooth 
topological density that could eventually be interpreted
in terms of continuum objects. 
%by certain solutions of the field equations. 
In the talk we have discussed and  
compared filtering via smearing, Laplacian eigenmodes and Dirac eigenmodes, as 
in \cite{Bruckmann:2006wf}. Preliminary results were presented in 
\cite{Gattringer:2006wq}. % Die DiGiacomo-Festschrift 
Our objective is to combine filtering methods in order to reduce ambiguities 
and highlight structures that are important at a certain level of smoothing.

Throughout this paper, we employ the \emph{highly improved field strength 
tensor} \cite{Bilson-Thompson:2002jk} which uses a 
weighted clover average of $1\times 1$, $2\times 2$ and $3\times3$-plaquettes. 
With its help a gluonic definition of the topological charge density can be 
given, that yields, for typical Monte Carlo configurations, integer topological 
charge with good precision after few smearing steps.
The configurations for zero temperature are $16^4$ lattices generated with 
tree-level improved L\"uscher-Weisz action at~\mbox{$\beta=1.95$.}

\section{Filtering techniques}

\subsection{Smearing}

\noindent
Smearing is a gauge covariant local averaging for the gauge field in order to 
remove UV fluctuations. Every gauge link is replaced by a weighted average of 
the link itself and the sum of the bypassing staples. The resulting average needs 
to be projected back to the gauge group. 
% In the case of the $SU(2)$ gauge 
% group, this projection is just dividing by real number. 
The following diagram,
sequentially applied to all links, visualizes one smearing step:
%%  muss das sein ? wenn ja, wuerde ich die Zeile "new link ..." aber
%%  weglassen.
% STEFAN:  nein, es muss nicht sein. Wenn wir noch Platz brauchen, können wir
% das auch weglassen  
\begin{fmffile}{staples}

\begin{center}

\begin{tabular}{ccccccc}
\begin{fmfgraph}(50,50)
\fmfstraight
\fmfleft{i,x}\fmfright{f,y}
\fmf{phantom}{i,ii}\fmf{phantom}{x,xx}\fmf{phantom}{f,ff}\fmf{phantom}{y,yy}
\fmf{phantom,tension=0.1}{ii,xx,yy,ff,ii}\fmffreeze
\fmf{double_arrow}{ii,ff}
\end{fmfgraph}
& 
$=$ 
&
$\alpha\;\times$
&
\begin{fmfgraph}(50,50)
\fmfstraight
\fmfleft{i,x}\fmfright{f,y}
\fmf{phantom}{i,ii}\fmf{phantom}{x,xx}\fmf{phantom}{f,ff}\fmf{phantom}{y,yy}
\fmf{phantom,tension=0.1}{ii,xx,yy,ff,ii}\fmffreeze
\fmf{fermion}{ii,ff}
\end{fmfgraph}
&
$+$
&
$\gamma\;\times$
&
\begin{fmfgraph}(50,50)
\fmfstraight
\fmfleft{i,x}\fmfright{f,y}
\fmf{phantom}{i,ii}\fmf{phantom}{x,xx}\fmf{phantom}{f,ff}\fmf{phantom}{y,yy}
\fmf{phantom,tension=0.1}{ii,xx,yy,ff,ii}\fmffreeze
\fmf{dots,tension=0}{ii,ff}
\fmf{fermion}{ii,xx,yy,ff}
\end{fmfgraph}\\
%
%% new link & $=$ & $\alpha\;\times$ & old link & $+$ & $\gamma\;\times$ & staples
\end{tabular}

\end{center}

\end{fmffile}  

\noindent
We used $\alpha=0.55$, $\gamma=0.075$ in all cases, following 
\cite{DeGrand:1997ss}. The smearing procedure is normally repeated several 
times to get stronger filtering, i.e., smoother configurations. The filtered 
topological charge is $q(x)$ computed from the smeared gauge fields using the 
improved field strength tensor:
\begin{gather}
	\label{eq:qsmeared}
    q^{\text{smeared}}(x)=
    \qhifmunu{(x)} 
    \quad\text{ with }
    \hif\text{ computed on smeared links.} 
    % STEFAN: das "\text... muss sein, sonst weiß man nicht, dass 
    %         der improved tensor gemeint ist
\end{gather}

\subsection{Laplace filtering}

\noindent 
The eigenmodes of the lattice Laplacian can 
%% also    wieso "also"  ?  
% STEFAN: also bezogen auf "smearing and Laplace"
also be used as a 
%% added "low-pass", um logsch an smearing anzuschliessen
% STEFAN: ich würde low-pass weglassen, denn der Begriff wird 
%         sonst nicht verwendet. 
low-pass filter for the gauge links \cite{Bruckmann:2005hy}.
%. ,Bruckmann:2005hy}.
% STEFAN: das hatte ich vorher drin, aber Falk meinte, eines reicht
This filter, however, is non-local. 
The lattice Laplacian 
\begin{gather}
    \label{eq:latticelaplacian}
    \Delta^{ab}_{xy}=
    \sum_{\mu=1}^{4}
    \left[
    U^{ab}_{\mu}(x) \delta_{x+\hat\mu,y}
    + U^{\dagger ab}_{\mu}(y) \delta_{x-\hat\mu,y}
   - 2\delta_{ab}\delta_{xy} \right]
\end{gather}
contains the original links, that can be reconstructed if all eigenmodes
of (\ref{eq:latticelaplacian}) are known.
For filtering, we truncate the sum in 
\begin{gather}
    \label{eq:laplacesum}
    U_\mu^{ab~\text{Laplace filtered}}(x) =
    -\sum_{n=1}^{N}
    \lambda_{n}\phi_n^a(x)
    \phi^{* b}_n(x+\hat\mu)\big|_{\text{normalization}}
\end{gather}
with $N\ll 2V$ with $V$ being the lattice volume. If $N=2V$, the 
formula is exact and the filtered gauge links would be identical to the 
original links. The topological charge is then computed via the improved 
field strength tensor on a configuration of Laplace filtered links, analogous 
to \eref{eq:qsmeared}.

\subsection{Dirac filtering}
 
\noindent
The third possibility to obtain a filtered topological charge density is to use 
eigenmodes of a chiral Dirac operator.  
We chose the \emph{chirally improved} (CI) Dirac operator $D^{CI}$
\cite{Gattringer:2000ja,Gattringer:2000js}. Its eigenmodes are reasonably
chiral, but the required computing time is much less than, e.g., for the overlap 
operator. The topological density $q(x)$ can be directly reconstructed from the 
eigenmodes, e.g.,  truncated to $N$ pairs of non-zero modes:
\begin{gather}
\label{eq:reconstructed}
\begin{aligned}
    q^{\text{rec.}}_N(x)
    &
    =
    \sum_{\substack{i=1\\ \lambda_i\text{ complex }}}^N\Bigg(
        \left(\frac{\lambda_i}{2}-1\right)
        \phi^\dagger_{\lambda_i}(x)\;
        \gamma_5\;\phi_{\lambda_i}(x)          
        +
        \left(\frac{\overline{\lambda_i}}{2}-1\right)
        \phi^\dagger_{\overline{\lambda}_i}(x)\;
        \gamma_5\;\phi_{\overline{\lambda}_i}(x)
        \Bigg)\\
    &
    +
    \sum_{\text{all real }\lambda_j}
    \left( - \frac{1}{|\rho_{5,\lambda_j}|}
             \phi^\dagger_{\lambda_j}(x)\;
                  \gamma_5\;\phi_{\lambda_j}(x)
    \right)
    \text{ with }
    \rho_{5,\lambda_j}=
    \sum_x\Big(\phi^\dagger_{\lambda_j}(x)\gamma_5\phi_{\lambda_j}(x)\Big)
\end{aligned}
\end{gather}
% STEFAN: das " with .... " soll in die Formel, spart eine Zeile 
The eigenmodes are defined as $D^{CI}\phi_n=\lambda_n\phi_n$, 
such that $\lambda$ is complex in general. 
The real modes determine the topological charge through the Atiyah-Singer index 
theorem. % STEFAN: das stimmt so 

Figure~\ref{fig:visualization} visualizes the filtering methods by displaying
the charge density over slices of the lattice. It shows a very good agreement of
the various methods. Similar observations have been made in  
\cite{DeGrand:2000gq}. 
%% Gehoert das hierher ?
%% STEFAN: Falk meine Ja. 
We want to emphasize that the agreement of the three methods is nontrivial since
the methods differ profoundly.  

\newcommand{\spx}{\rule{0pt}{4pt}}

\begin{figure}[t!]
\begin{center}
\begin{tabular}{b{0.2\textwidth}|b{0.2\textwidth}b{0.2\textwidth}b{0.2\textwidth}}
\includegraphics[width=0.2\textwidth]{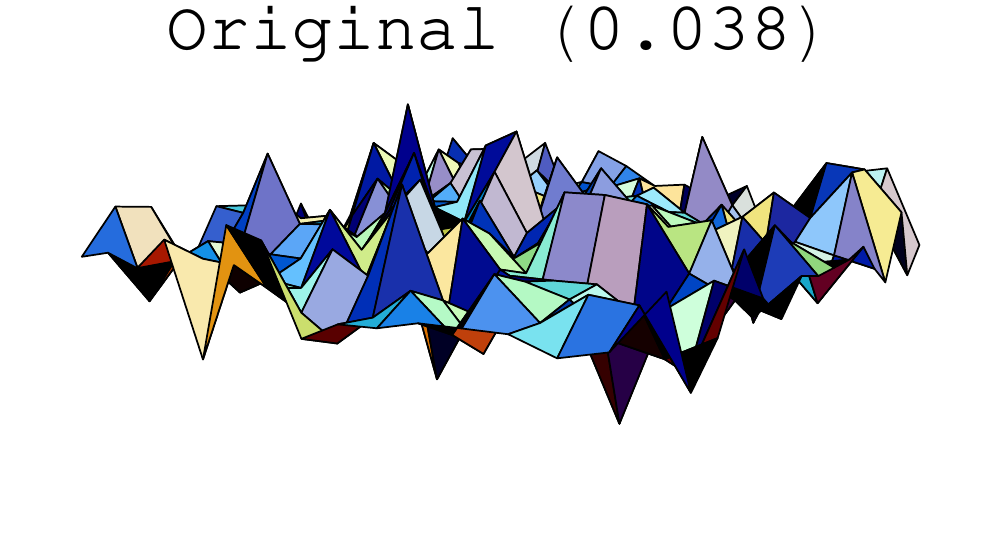}&
\includegraphics[width=0.2\textwidth]{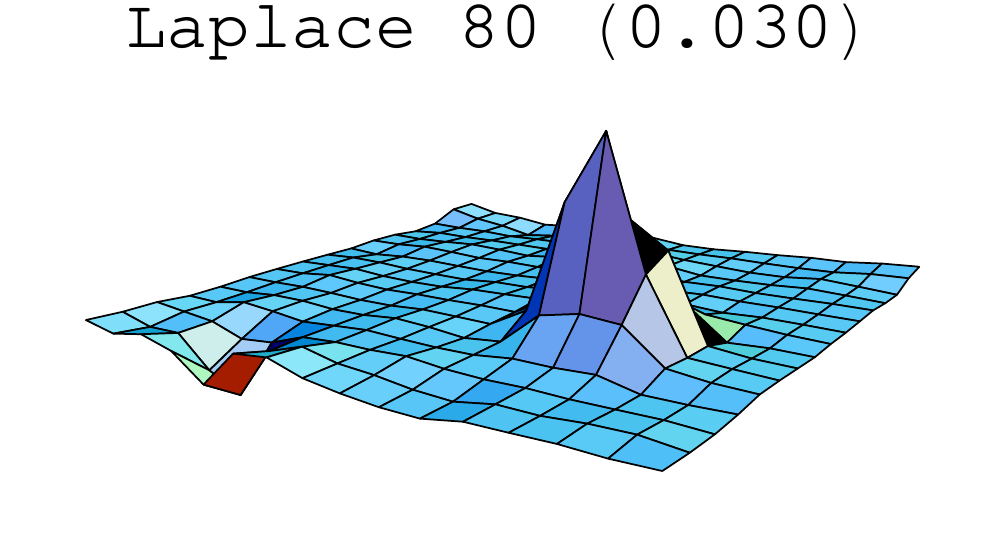}&
\includegraphics[width=0.2\textwidth]{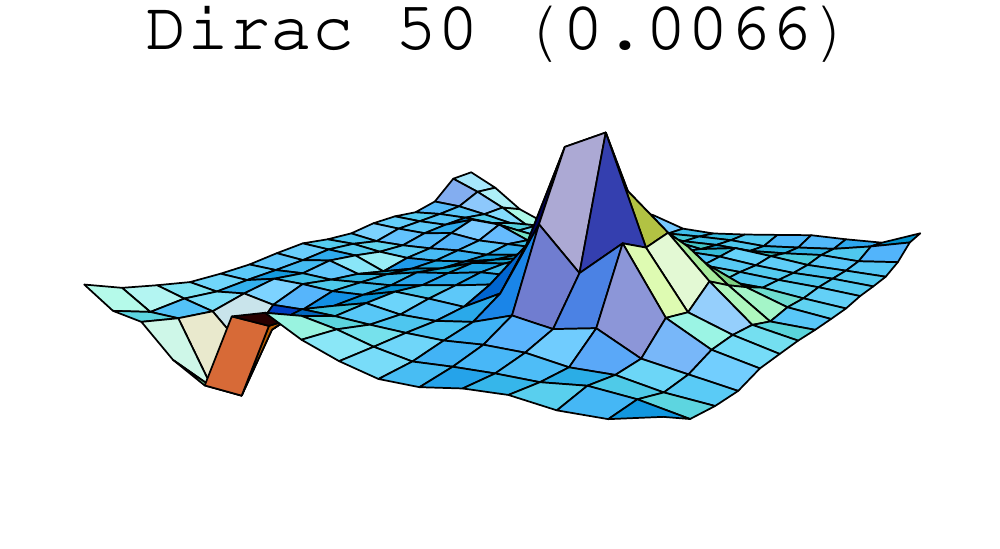}&
\includegraphics[width=0.2\textwidth]{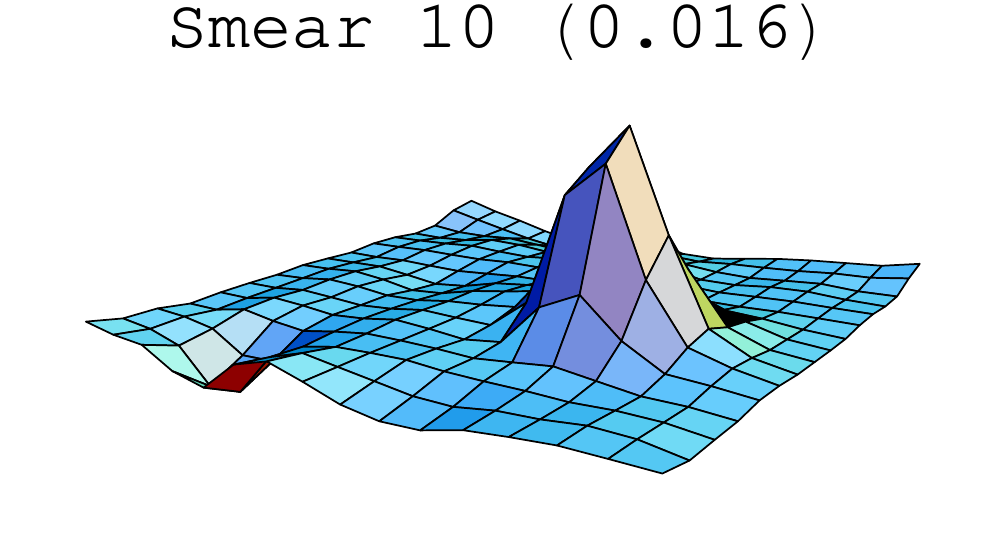}\\ \hline
\spx\par\includegraphics[width=0.2\textwidth]{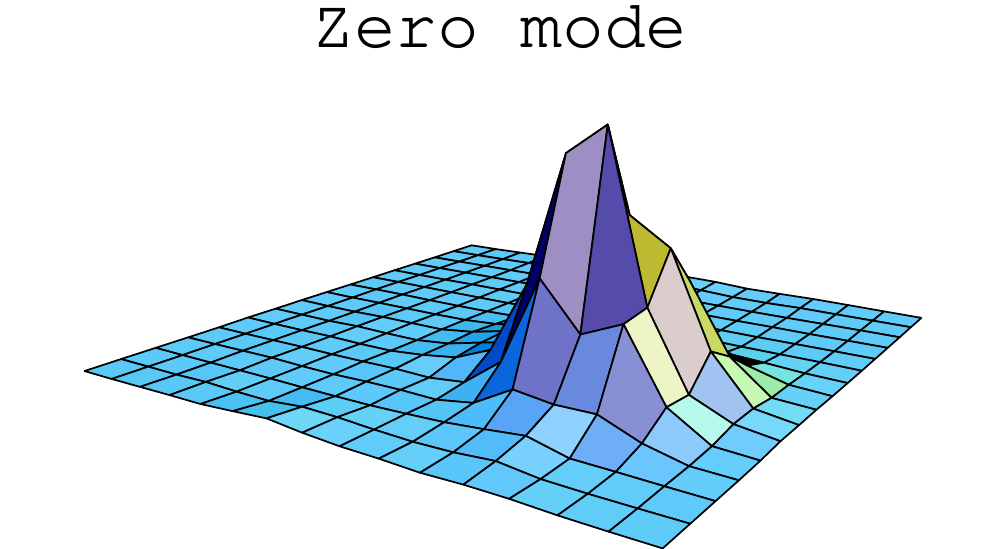}&
\spx\par\includegraphics[width=0.2\textwidth]{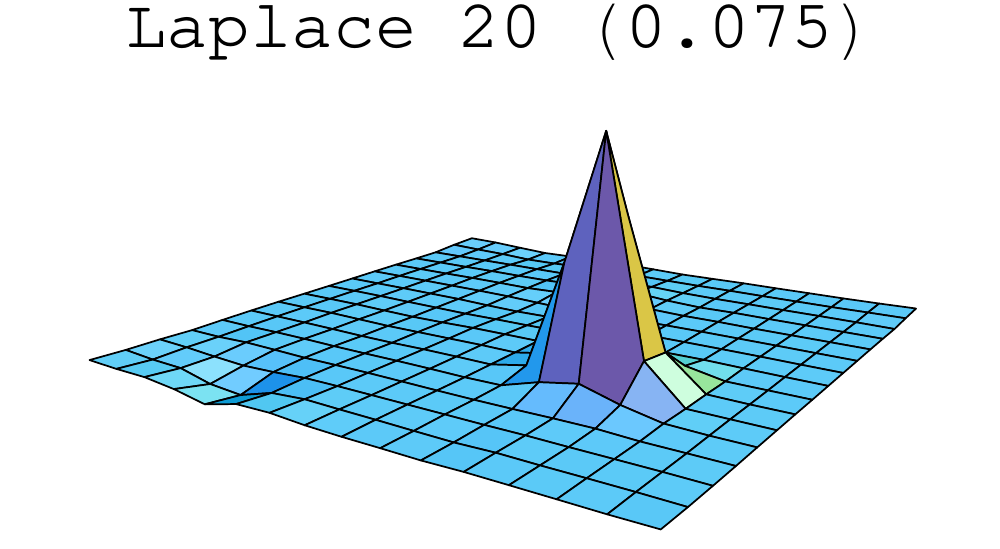}&
\spx\par\includegraphics[width=0.2\textwidth]{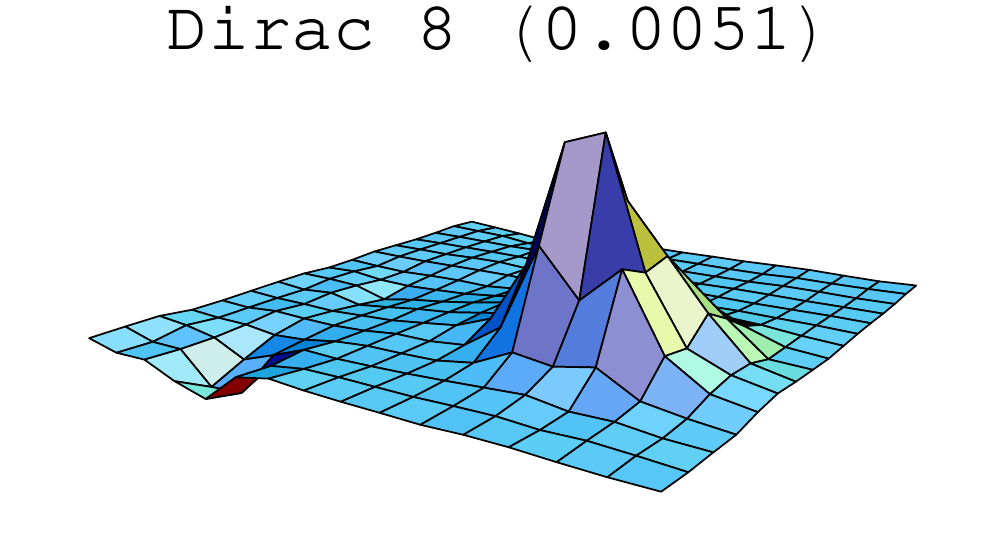}&
\spx\par\includegraphics[width=0.2\textwidth]{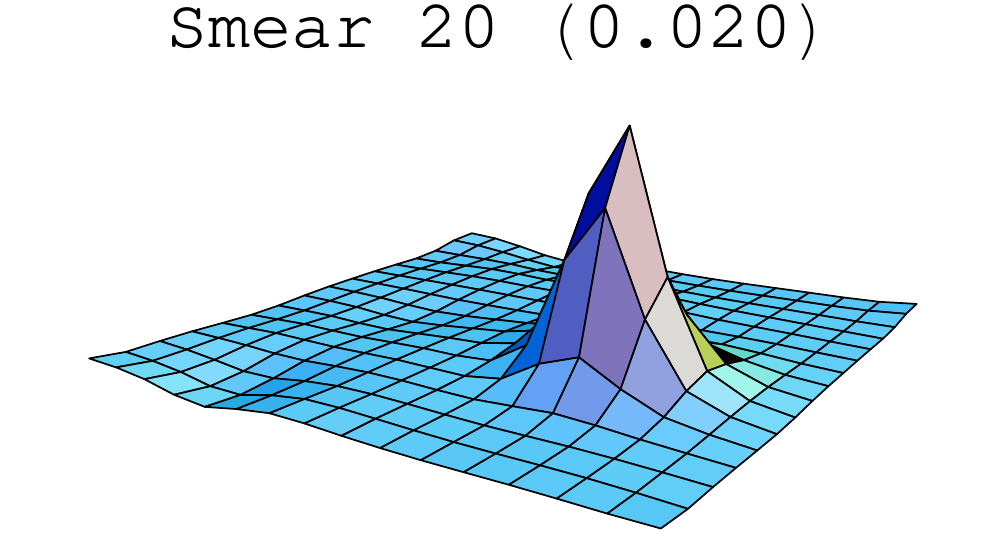}
\end{tabular}
\end{center}
\caption[Visualization of filtering]{The plots visualize the filters. All 
images show the same 2-dimensional slice of the topological charge density of 
the same configuration. The upper left plot shows the unfiltered density. The 
lower left plot shows  the scalar density of the single real eigenmode for 
comparison. The other plots show the filtered topological charge with the three 
methods and various filtering parameters. Note that the plots are drawn
not in the same scale. Peak values are indicated in brackets.}
\label{fig:visualization}
\end{figure}

\section{Optimal filtering}

\noindent
Optimal filtering is a way to match the 
three filter parameters -- the number of Dirac modes, of Laplace modes and  
of smearing steps -- in such a way, that the resulting topological 
charge density obtained by either of the methods resembles the topological charge 
density obtained by the other methods. 
%% muss nicht erklaert werden, was $\chi$ ist ?
% STEFAN: wird unten erklaert
We consider the correlators and cross correlators $\chi_{q_A q_B}(r)$ of the
topological charge densities $q_A$, $q_B$ filtered with filter $A$ and $B$. 
%% Eigentlich q_A(x) - \bar{q}_A !
% STEFAN: ja, eigentlich schon. macht aber hier die 
%         Formel unnoetig lang und aendert am Ergebnis nichts
Then the relative normalization factor 
%% was waere das fuer 0 --> r beliebig ?
% STEFAN: einfach "0" durch "r" ersetzen. 
%         dann sind es aber auch die \chi(r) und 
%         eine anschauliche Interpretation wird schwerer
%%  $\Xi_{q_A q_B}(0)$  
\begin{gather}
\label{eq:Xi}
\Xi_{q_A q_B}\equiv
    \frac
    {\big< \chi_{q_A q_B}(0) \big>\;\big< \chi_{q_A q_B}(0) \big>}
    {\big< \chi_{q_A q_A}(0) \big>\;\big< \chi_{q_B q_B}(0) \big>} 
    \quad\text{ with }\quad
    \chi_{q_a\;q_B}(0)\equiv(1/V)\sum_x\;q_A(x)\;q_B(x)  
\end{gather}
is a measure of the similarity of the topological charge densities $q_A$, $q_B$.
Two filtered configurations are the more similar, the closer $\Xi$ is~to~$1$. 

\begin{figure}[t!]
\begin{center}
\begin{tabular}{ll}
\includegraphics[width=0.4\textwidth]{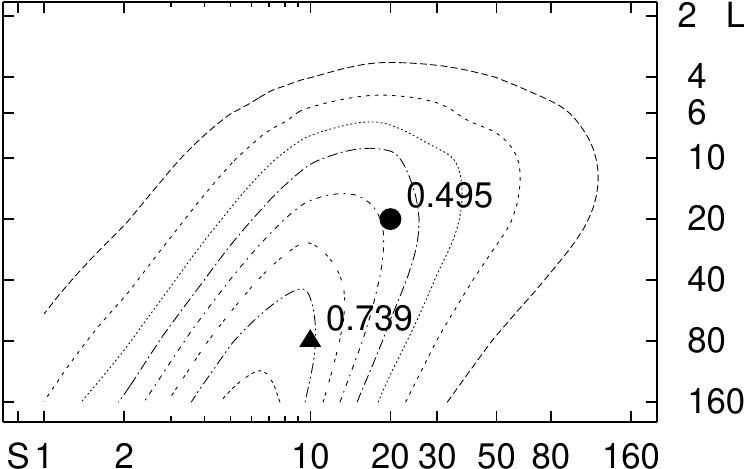}&
\includegraphics[width=0.4\textwidth]{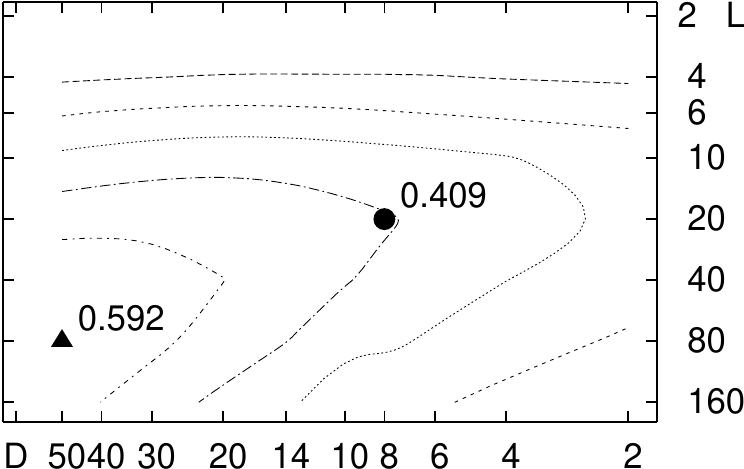}\\
% \begin{minipage}[b]{0.45\textwidth}
% \begin{gather*}
%         \text{lines of constant }\Xi\\
%     \Xi_{q_A q_B}(0)=\\
%     \frac
%     {\big< \chi_{q_A q_B}(0) \big>\;\big< \chi_{q_A q_B} (0)\big>}
%     {\big<\chi_{q_A q_A}(0)\big>\;\big<\chi_{q_B q_B}(0)\big>}
% \end{gather*}
% \end{minipage}
&
\includegraphics[width=0.4\textwidth]{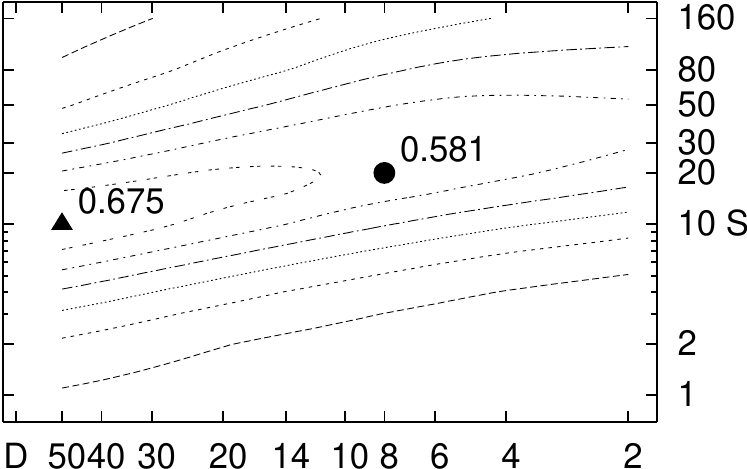}
\end{tabular}
\end{center}
% \caption[Optimal filtering]{The above plots show how to determine optimal
% filtering. The lines in the plots denote constant values of $\Xi$. The
% triangle $\blacktriangle$ denotes the position of weak filtering, the dot
% {\Large$\bullet$} denotes the position of strong filtering. Say we choose an
% arbitrary number of Laplace modes used for filtering, e.g., 20 modes. If we
% want to know what number of Dirac modes corresponds to 20 Laplace modes, I
% look at the upper right plot. The lines of constant $\Xi$ form a ``ridge.'' 
% The point where 20 Laplace modes lie on the ridge, gives the number of
% corresponding Dirac modes. Here, it is 8 Dirac modes, marked with a dot 
% {\Large$\bullet$}. In the same fashion, I determine from the lower right plot
% that 8 Dirac modes correspond to 20 smearing steps. As a cross check, one can
% look at the left plot. The position `` 20 Laplace modes, 20 smearing steps''
% is very close to the ridge. This tells us that the method to match filtering
% parameters is consistent between all three methods. }
\caption[Optimal filtering]{
% STEFAN: ich mag lieber ganze Saetze. 
These plots show a pairwise comparison of filtering methods. 
Shown are lines of constant $\Xi$, as defined in \eref{eq:Xi}.
L, D and S refer to Laplace and Dirac filtering and smearing.
The ridges in each contour plot represent the optimal matching
between two methods. The triangle $\blacktriangle$ denotes 
an example of weak filtering, the dot {\Large$\bullet$} 
represents a case of strong filtering in all three methods. 
The dots in the right plots relate 20 Laplacian modes to 8 pairs of non-zero 
Dirac modes, and these to 20 smearing steps. The dot in the left plot is
close to the ridge there, confirming the consistency of matching between 
all three methods.}
% STEFAN: das folgende will ich schon mit dabei haben.
%         dieser Prozess war schon beim Vortrag nicht allen ganz klar, 
%         und ich moechte das lieber ausfuehrlich 
% Say we choose an arbitrary number of Laplace modes used for filtering, e.g., 20 
% modes. Then the ridge in the upper right plot gives the corresponding number of 
% Dirac modes, here 8, marked with the dot {\Large$\bullet$}. In the same 
% fashion, we determine from the lower right plot that 8 Dirac modes correspond 
% to 20 smearing steps. As a cross check, the position ``20 Laplace modes, 20 
% smearing steps'' in the left plot is very close to the ridge. This confirms 
% that the method to match filtering parameters is consistent between all three 
% methods. }
\label{fig:optimalfiltering}
\end{figure}

\section{Clusters}

\noindent Once we have computed the topological charge density, we can try to 
identify individual objects. It is well known that the topological charge in 
smoothed configurations shows a ``lumpy'' structure.

We characterize a certain cluster state in different lattice configurations 
using a running ``watermark'' $f$ rather than using a cutoff $q_{cut}$. This 
means that the lower cutoff $q_{cut}$ in $|q(x)|$ for sites assigned to any 
cluster in a given configuration is adapted according to the resulting packing 
fraction $f$. Compared to considering cluster states being equivalent at equal 
$q_{cut}$, the watermark method reduces the noise from configuration to 
configuration. 
A \emph{cluster} is then defined as 
a connected set of lattice sites lying above the watermark and having the
same sign of the topological charge density.
Choosing $f$ we always have approximately the same number of 
clusters in all configurations. If we would compare configurations at fixed 
cutoff $q_{cut}$, there would be configurations with no clusters as well as 
configurations with
many clusters.

\begin{figure}[t!]
\begin{center}
  \includegraphics[width=0.48\textwidth]{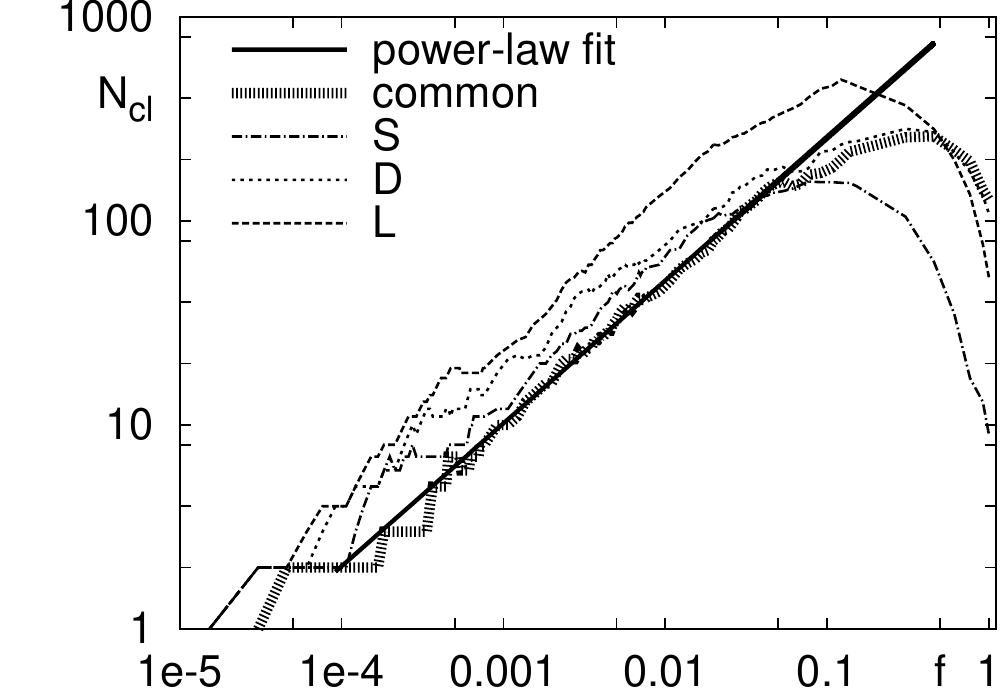}
  \hfill
  \includegraphics[width=0.48\textwidth,clip=false]{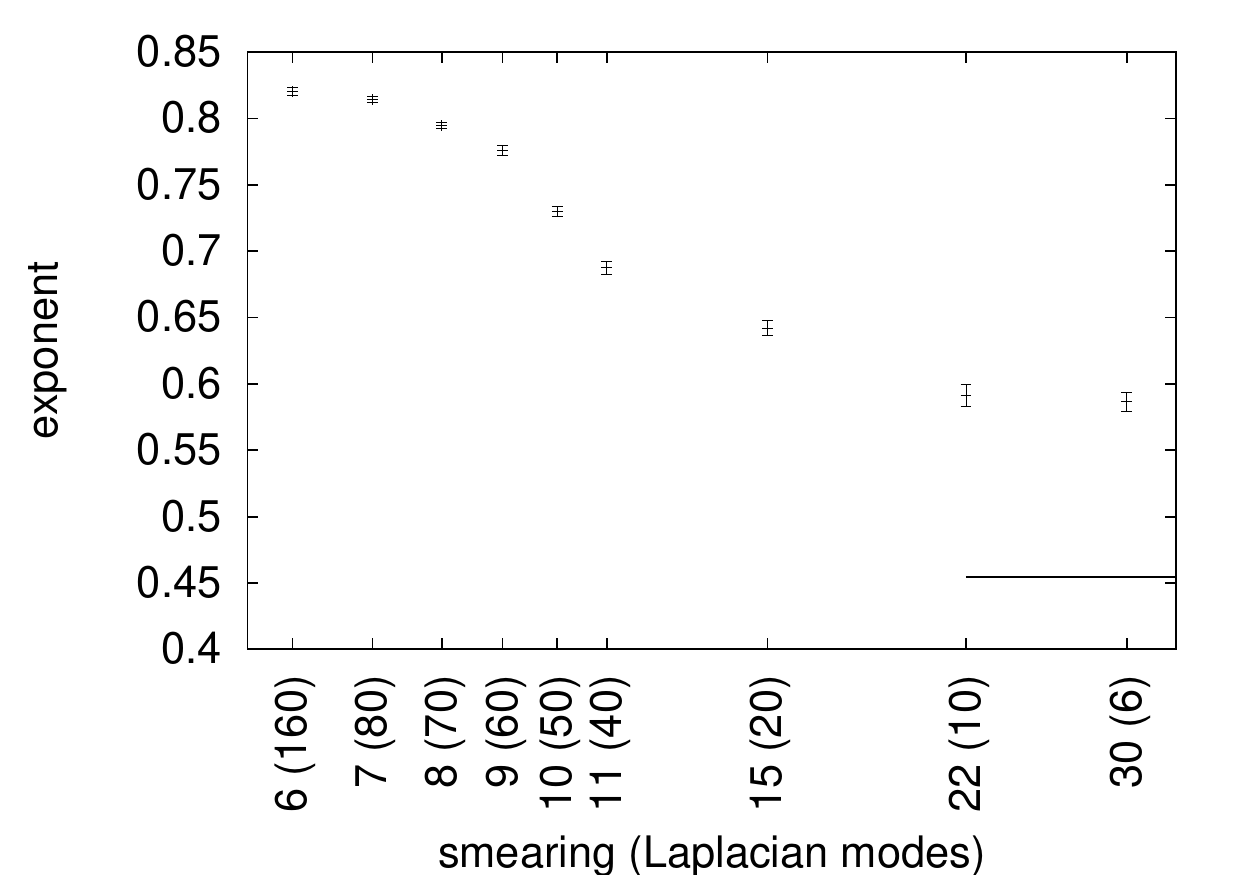}
  \caption[watermark]{The left, double-log plot shows the number of clusters 
  vs.\ the packing fraction $f$ for various filtering methods. Note that for 
  the thick, dotted line (labelled ``common'') the linear regime is wider than 
  for the individual filtering methods.
%% What is "common" is this explained in the text ? STEFAN: ja  
  This plot uses weak filtering, i.e., 10 smearing steps, 80 Laplace modes and 
  50 Dirac modes. The right hand side plot shows the exponent of the power law, 
  obtained when only clusters are taken into account found by both smearing and 
  Laplace filtering. The abscissa of the r.h.s.\ plot describes the degree of 
  smoothing by the number of smearing steps (in parentheses the corresponding 
  number of Laplace modes).}
  \label{fig:watermark}
\end{center}
\end{figure}

%%\subsection{A power law}

%\enlargethispage*{1\baselineskip}

\noindent
Figure~\ref{fig:watermark} (left) shows the changing cluster composition for 
various filtering methods, as function of $f$. In particular, the method 
denoted as \emph{common} is interesting: Only those points are considered that 
are found by all three filtering methods. All curves show a pronounced power 
law for a small packing fraction $f$.
% STEFAN: Hier kam der Wunsch auf, "percolation" mit einzubringen: 
A decrease of the curves signals the beginning of cluster percolation. 

The remaining question is: does the exponent of the power law depend on the
level of filtering?  
%% Kam das unerwartet ?
% STEFAN: Ja, schon. Ich haette urspruenglich erwartet, 
% dass es nicht so stark schwankt.  
We found that the exponent of the power law describes a cluster composition
that depends on the level of filtering.
The right hand side plot of Figure~\ref{fig:watermark} demonstrates this.
The exponent indeed depends on the level of filtering. It changes most rapidly
around 10 smearing steps, but reaches a plateau around 30. 
Data points for more than 30 smearing steps -- or less than 6 Laplacian modes --
do not make sense in this analysis since there would be so few clusters per 
configuration that we can hardly speak of a homogeneous cluster state.
%% Wie gut ist in diesem Lichte das Plateau ?
% STEFAN: schon recht gut. es bricht nicht ein, solange man 
%         genuegent cluster hat. 
The horizontal mark in the right hand side plot in Figure~\ref{fig:watermark} 
denotes the exponent that would correspond to the dilute instanton gas. From 
that plot, we can clearly see that the dilute instanton gas as a model for 
topological objects is excluded by our data.

\section{Finite Temperature}

\noindent 
Up to now, all results described zero temperature. 
Here we extend our investigation to the behavior of the topological lumps on 
both sides of the phase transition.

We used several ensembles above and below the critical temperature. 
Our set of ensembles consists of configurations with the same inverse
coupling $\beta=1.95$, 
with a spatial extent $N_s^3=20^3$ of the lattice and
varying temporal size, $N_t=4,6,8,10,12$. Each ensemble consists of 100
thermalized configurations.
% Figure~\ref{fig:efinitepolyakovcorrelator} shows
% the static quark potential computed from Polyakov correlators for these
% ensembles. 
The ensembles with $N_t=4,6$ are clearly above the phase transition,
$N_t=10,12$ are below the phase transition, and $N_t=8$ is roughly at the
phase transition. 

% \begin{figure}[phtb]
% \begin{center}
%   % The graphic below is the output of
%   % /home/sos28559/workspace/MaxAbelianGauge/plot_potential_v_from_PL.gnu
%   % converted to PDF with epstopdf. 
%   \includegraphics[width=0.5\textwidth]{efinitepolyakovcorrelator.pdf}
%   \caption[Potential from PL correlators for varying $N_T$]{The plots show the
%   static quark potential computed via Polyakov loop correlators. The lines are
%   fitted functions of the form $c+b/R+SR$. The numbers after the word
%   \emph{tension} are $\sqrt{S}$, including the errors. The data points for a
%   given ensemble have been shifted by an arbitrary value such that they all
%   fit in one graph.}
%   \label{fig:efinitepolyakovcorrelator}
% \end{center}
% \end{figure}

%\subsection{Results}

\enlargethispage*{2\baselineskip}

\begin{figure}[t!]
\begin{center}
  %\begin{minipage}[b]{0.34\textwidth}
  %\includegraphics[width=0.48\textwidth]{pl-in-clusters-high-T.pdf}
  %\hfill
  %\includegraphics[width=0.48\textwidth]{pl-in-clusters-low-T.pdf}
  
  \vspace{-1\baselineskip}
  
  % diese Dateien sind aus: 
  % /misc/kernnas2/sos28559/selfgen/finite/b1950/ploops/lattice-plots.ps
  % seiten 1,4,5,8 
  % erzeugt mit lattice-plots.sh 
  % 
  \includegraphics[width=0.49\textwidth,
  bb= 0 0 600 400]{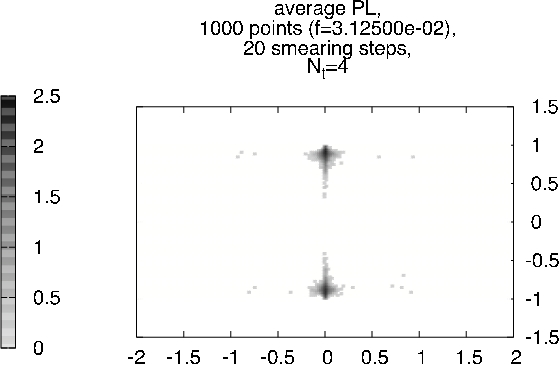} \hfill
  \includegraphics[width=0.49\textwidth,
  bb= 0 0 600 400]{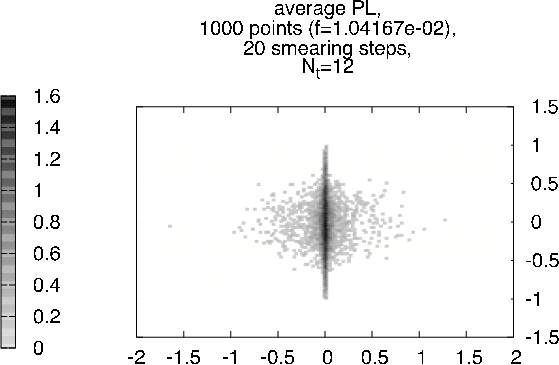}
  
  \includegraphics[width=0.49\textwidth,
  bb= 0 0 600 400]{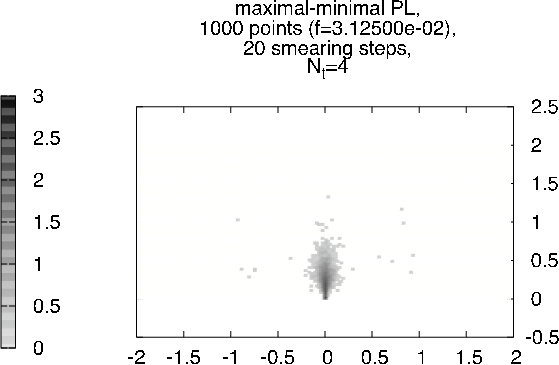} \hfill
  \includegraphics[width=0.49\textwidth,
  bb= 0 0 600 400]{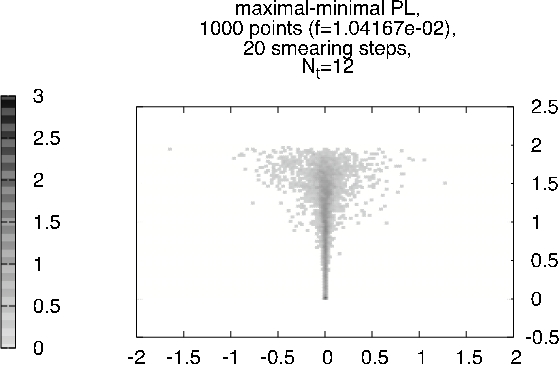}
  
  %\end{minipage}
  %\end{minipage}
  %\hfill
  %\vrule width 1pt\relax
  
%%   die plot-Beschriftung ragt ziemlich ueber.
%   STEFAN: wie ist das gemeint?  
%   \vspace{\baselineskip}
  \includegraphics[width=0.4\textwidth]{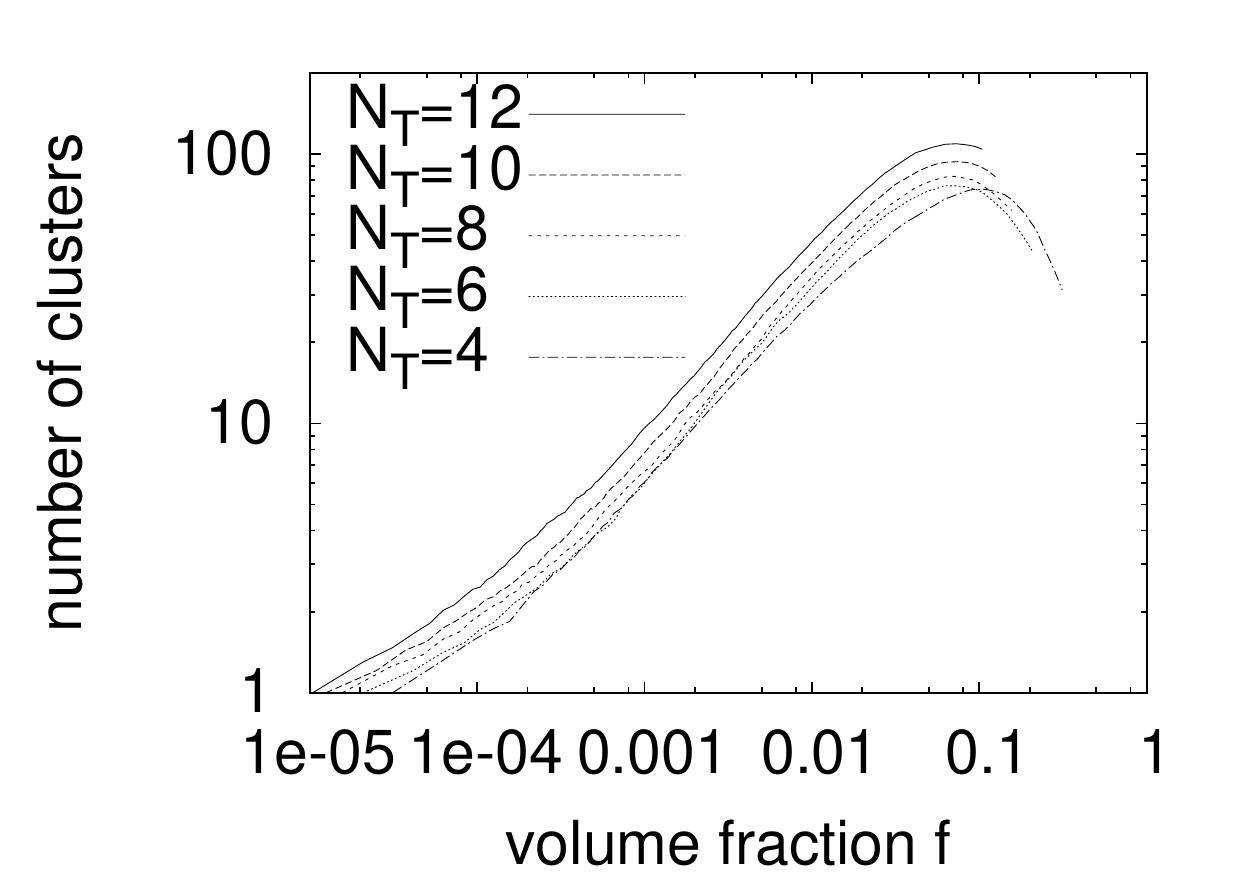}
  \caption[Cluster distribution for finite $T$]
  {At the top, we show 2D-histograms of topological lumps. The abscissa 
  indicates the total topological charge inside the cluster, the ordinate shows 
  the averaged Polyakov loop through the cluster and the ``grayness'' shows the 
  number of lumps in a given bin. The clusters indeed feel the phase 
  transition.
  The middle row is similar to the top row, but shows the difference between the
  maximal and the minimal Polyakov loop within a cluster.
  At the bottom, we show in a double-log plot the number of clusters vs. 
  packing fraction for finite
  temperature. The absolute number of the clusters 
  varies, but this might be due to the varying total lattice volume.
%% Ist es nun proportional oder nicht ?
% STEFAN: grob. nicht ganz proportional. 
%         koennten auf finite-size effekte sein  
  The slope of the distribution does apparently not change.
  }
%% aber der lineare Bereich ist viel kuerzer.  
% STEFAN: finde ich nicht so dramatisch. 
  \label{fig:powerlawsfinite}
\end{center}
\end{figure}

The upper plots in Figure~\ref{fig:powerlawsfinite} show 2D-histograms of 
clusters. The abscissa shows the topological charge inside a given cluster, the 
ordinate shows the Polyakov loop averaged within the same cluster. The gray 
value indicates the height of a given bin. The top left plot is for 20 smearing 
steps and $N_t=4$, the top right plot is for 20 smearing steps and $N_t=12$. 
Clearly, the clusters themselves \emph{do} feel the phase transition. The 
Polyakov loop inside a given cluster reflects the distribution of the global 
average of the Polyakov loop: in the deconfined phase the distribution is 
strongly peaked close to $L=\pm1$, in the confined phase the distribution is
wide, centered at $0$. 
%\enlargethispage*{\baselineskip}

The middle row of Figure~\ref{fig:powerlawsfinite} also shows 2D-histograms just
like the top row, but the
difference between the minimal and the maximal Polyakov loop within a cluster is
shown. 
Above the phase transition this difference is small but below the phase
transition larger clusters reveal a dipole-like structure in the local Polyakov
loop. See also~\cite{Ilgenfritz:2006ju}.

%% Aber ruecken die f's nicht zu kleineren Werten und ebenso die cluster-Zahlen ?
%% Was aendert sich also nicht ?
%% Ist der lineare Bereich nicht auch viel kleiner ?
% STEFAN: der ist nur ein bisschen kleiner. Das kann man alles damit
%         erklaeren, dass einfach das Volumen kleiner ist. 
The lowest panel of Figure~\ref{fig:powerlawsfinite} shows the number of
clusters  vs.\ $f$ for the five ensembles with $N_t=4,...,12$ at $\beta=1.95$. 
We see that varying $N_t$, such that the ensembles are below 
or above the phase transition, seems to have no effect on the cluster 
composition as a function of $f$.

\section{Conclusions}

\noindent
We conclude that all three filtering methods corroborate each other. If their 
parameters are mapped onto each other, they yield similar results. Especially 
smearing has often been criticized for producing arbitrary results. However, 
our results show -- for moderate smearing with 10 to 30 smearing steps --
%%that smearing is capable to mimick
% STEFAN: "mimick" finde ich zu negativ 
that smearing keeps up with 
%% not better or worse than 
the other filtering methods. For this regime, the power law we found excludes 
the dilute instanton gas as a model for topological structures. Surprisingly, the 
power law survives the phase transition almost unchanged. This peculiar 
behavior requires further studies. The internal characteristics of the clusters 
--- topological charge and Polyakov loop ---  strongly change with the onset of 
deconfinement. 
    
\bibliography{disbib}

\end{document}